\documentclass[12pt]{article}
\usepackage{amssym}
\usepackage{graphics}

\font\elevenof=msbm10 at 11pt
\font\sixteenof=msbm10 at 16pt

\def\C{\mbox{$\Bbb C$}}
\def\R{\mbox{$\Bbb R$}}
\def\Qp{Q^{\dagger}}
\def\Ap{A^{\dagger}}
\def\case#1#2{{\textstyle{#1\over #2}}}
\def\sech{\mathop{\rm sech}\nolimits}
\def\cosech{\mathop{\rm cosech}\nolimits}

\title{
\hfill{\normalsize ULB/229/CQ/00/7}\\
\vspace{1cm} 
GENERATING COMPLEX POTENTIALS WITH REAL EIGENVALUES IN SUPERSYMMETRIC
QUANTUM MECHANICS}
\author{B. BAGCHI\thanks{E-mail: bbagchi@cucc.ernet.in} \ and S.
MALLIK \\
{\small \sl Department of Applied Mathematics, University of Calcutta,} \\
{\small \sl 92 Acharya Prafulla Chandra Road, Calcutta 700 009, India}\\[10pt] 
C. QUESNE\thanks{Directeur de recherches FNRS; E-mail: cquesne@ulb.ac.be} \\
{\small \sl Physique Nucl\'eaire Th\'eorique et Physique Math\'ematique,}\\ {\small \sl
Universit\'e Libre de Bruxelles, Campus de la Plaine CP229,} \\ {\small \sl  Boulevard~du
Triomphe, B-1050 Brussels, Belgium}}
\date{ }
\begin{document}
\baselineskip=22pt plus 1pt minus 1pt
%%%%%%%%%%%%%%%%%%%%%%%%%%%%%%%%%%%%%%%%%%%%%%%%%%%%%%%%%%
\maketitle

\begin{abstract} 
In the framework of SUSYQM extended to deal with non-Hermitian Hamiltonians, we
analyze three sets of complex potentials with real spectra, recently derived by a potential
algebraic approach based upon the complex Lie algebra sl(2,\mbox{\elevenof C}). This
extends to the complex domain the well-known relationship between SUSYQM and potential
algebras for Hermitian Hamiltonians, resulting from their common link with the factorization
method and Darboux transformations. In the same framework, we also generate for the
first time a pair of elliptic partner potentials of Weierstrass $\wp$ type, one of them being
real and the other imaginary and PT symmetric. The latter turns out to be quasiexactly
solvable with one known eigenvalue corresponding to a bound state. When the
Weierstrass function degenerates to a hyperbolic one, the imaginary potential becomes
PT non-symmetric and its known eigenvalue corresponds to an unbound state.
\end{abstract}

\vspace{1cm}
\noindent
Running head: Complex Potentials in SUSYQM

\newpage
%
%========================================================================
%
\section{Introduction}
Recently there has been some interest~\cite{bender, andrianov, znojil, bagchi00a,
bagchi00b, bagchi00c,khare} in studying PT-symmetric quantum mechanical systems. In
quantum mechanics, the Hamiltonian of the underlying system is usually assumed
Hermitian ensuring a real energy spectrum. However it has been
conjectured~\cite{bender} that under less restrictive situations, namely by requiring the
Schr\"odinger Hamiltonian to be invariant under the joint action of parity (P) and time
reversal (T) transformations, one can still have a real spectrum of energy eigenvalues.
Moreover, the overall normalizability of wave functions in many cases is not affected. In the
literature, PT-symmetric schemes have been explored with respect to the complexification
of several well-known potentials. Further, new ones have been searched for using a variety
of techniques~\cite{znojil, bagchi00a, bagchi00b,bagchi00c,khare}.\par
%
%-------------------------------------------------------------------------------------------------------------
%
In this paper, we shall present results from a supersymmetric point of view. We shall
show that the constraints furnished by the commutation relations of the sl(2,\C) algebra
admit of solutions that are consistent with supersymmetric intertwining relations. We
shall also exploit the latter to obtain elliptic solutions of Weierstrass $\wp$ type. Indeed
we shall report here for the first time a PT-symmetric potential defined in terms of 
Weierstrass $\wp$ function.\par
%
%=============================================================
% 
\section{\boldmath $N=2$ SUSYQM and Its Complexification Procedure}
\setcounter{equation}{0}

In order to set the notations, it would be useful to briefly recall the key features of the
one-dimensional $N=2$ SUSY quantum mechanics (SUSYQM). As is well
known~\cite{cooper}, the latter involves a pair of supercharges $Q$ and $\Qp$, related
by Hermitian conjugation, and in terms of which the governing Hamiltonian $H_s$ is
expressed as
\begin{equation}
  H_s = \left\{Q, \Qp\right\}.
\end{equation}
The supercharges $Q$ and $\Qp$ are fermionic in character and commute with $H_s$:
\begin{eqnarray}
  Q^2 & = & \left(\Qp\right)^2 = 0, \nonumber \\[0pt]
  [Q, H_s] & = & \left[\Qp, H_s\right] = 0.
\end{eqnarray}
\par
%
%-----------------------------------------------------------------------------------------------------------
%
A convenient way to deal with $Q$ and $\Qp$ is to adopt the representations
\begin{equation}
  Q = A \otimes \sigma_-, \qquad \Qp = \Ap \otimes \sigma_+, 
\end{equation}
where $A$ is some linear differential operator and $\sigma_{\pm}$ are combinations
$\sigma_{\pm} = \sigma_1 \pm {\rm i} \sigma_2$ of the Pauli matrices. A
first-derivative realization of $A$ and $A^{\dagger}$, namely
\begin{equation}
  A = \frac{d}{dx} + W(x), \qquad \Ap = - \frac{d}{dx} + W(x),
\end{equation}
yields the forms
\begin{equation}
  Q = \left(\begin{array}{cc}
         0 & 0 \\
         \frac{d}{dx} + W & 0
         \end{array}\right), \qquad
  \Qp = \left(\begin{array}{cc}
           0 & - \frac{d}{dx} + W \\
           0 & 0
         \end{array}\right),  \label{eq:supercharges}
\end{equation}
where $W(x)$ is the so-called superpotential of the system. Note that
(\ref{eq:supercharges}) renders $H_s$ diagonal,
\begin{equation}
  H_s = \left(\begin{array}{cc}
            H_+ & 0 \\
            0 & H_-
            \end{array}\right).
\end{equation}
As such we can factorize $H_{\pm}$ in the manner
\begin{eqnarray}
  H_+ & = & \Ap A = - \frac{d^2}{dx^2} + V^{(+)}(x) - E, \nonumber \\
  H_- & = & A \Ap = - \frac{d^2}{dx^2} + V^{(-)}(x) - E,  \label{eq:H_+H_-}
\end{eqnarray}
at some arbitrary factorization energy $E$ , with the partner potentials $V^{(\pm)}$
related to $W(x)$ through
\begin{equation}
  V^{(\pm)} = W^2 \mp W' + E.  \label{eq:Vpm}
\end{equation}
\par
%
%----------------------------------------------------------------------------------------------------
% 
It is easy to be convinced that the spectra of $H_+$ and $H_-$ are alike except possibly
for the ground state. In the exact SUSY case to which we shall restrict ourselves here, the
ground state at vanishing energy is nondegenerate and is associated with $H_+$ or
$H_-$,
\begin{equation}
  H_+ \psi^{(+)}_0(x) = 0, \qquad \psi^{(+)}_0(x) = K \exp \left(- \int^x W(t)
  dt\right),  \label{eq:gs+} 
\end{equation}
or
\begin{equation}
  H_- \psi^{(-)}_0(x) = 0, \qquad \psi^{(-))}_0(x) = K \exp \left(\int^x W(t)
          dt\right),  \label{eq:gs-}
\end{equation}
according to whether $\int^x W(t) dt \to + \infty$ or $- \infty$ as $x \to \pm \infty$.
In~(\ref{eq:gs+}) and~(\ref{eq:gs-}), $K$ is some normalization constant.\par
%
%-----------------------------------------------------------------------------------------------------------
%
The double degeneracy of the spectrum of $H_s$ can also be summarized by intertwining
$H_+$ and $H_-$ according to
\begin{equation}
  A H_+ = H_- A, \qquad H_+ \Ap = \Ap H_-.
\end{equation}
These relations follow from~(\ref{eq:H_+H_-}).\par
%
%-----------------------------------------------------------------------------------------------------------
%
To generate non-Hermitian potentials within SUSYQM~\cite{andrianov}, it is instructive to
decompose the underlying superpotential $W(x)$, the partner potentials $V^{(\pm)}(x)$,
and the factorization energy $E$ into a real and an imaginary part, namely
\begin{eqnarray}
  W(x) & = & f(x) + {\rm i} g(x),  \label{eq:W} \\
  V^{(+)}(x) & = & V^{(+)}_R(x) + {\rm i} V^{(+)}_I(x), \\
  V^{(-)}(x) & = & V^{(-)}_R(x) + {\rm i} V^{(-)}_I(x), \\
  E & = & E_R + {\rm i} E_I  \label{eq:E},   
\end{eqnarray}
where $f$, $g$, $V^{(\pm)}_R$, $V^{(\pm)}_I$, $E_R$, and $E_I \in \R$. All this leads
to SUSY without Hermiticity: in particular, the supercharges are no longer related by
Hermitian conjugation. As will be evident below, this presents no difficulty in so far as
developing a theoretical framework is concerned.\par
%
%---------------------------------------------------------------------------------------------------------------
%
{}From (\ref{eq:Vpm}) and (\ref{eq:W})--(\ref{eq:E}), it follows that
\begin{eqnarray}
  V^{(+)}_R & = & f^2 - g^2 - f' + E_R, \\
  V^{(+)}_I & = & 2fg - g' + E_I, \\
  V^{(-)}_R & = & f^2 - g^2 + f' + E_R, \\
  V^{(-)}_I & = & 2fg + g' + E_I.
\end{eqnarray}
These expressions are consistent with intertwining relationships.\par
%
%-----------------------------------------------------------------------------------------------------------
%
Since we will be interested only in a real energy spectrum, we can set $E_I = 0$. As such
our basic relations correspond to
\begin{eqnarray}
  V^{(+)}_R & = & f^2 - g^2 - f' + E_R,  \label{eq:V^{(+)}_R} \\
  V^{(+)}_I & = & 2fg - g', \\
  V^{(-)}_R & = & f^2 - g^2 + f' + E_R, \\
  V^{(-)}_I & = & 2fg + g'.  \label{eq:V^{(-)}_I}
\end{eqnarray}
Observe that $V^{(+)}_R$ and $V^{(-)}_R$ are related by $f \to -f$, while
$V^{(+)}_I$ and $V^{(-)}_I$ are linked through $g \to -g$. In the following our task will
be to analyze plausible solutions for the functions $f$ and $g$ pertaining to the set
(\ref{eq:V^{(+)}_R})--(\ref{eq:V^{(-)}_I}), including the PT-symmetric ones. The main
point to be noted is that by imposing an additional PT-symmetric restriction the Hermitian
character of the intertwined Hamiltonians is lost.\par
%
%----------------------------------------------------------------------------------------------------------
%
We shall first of all seek connections with the sl(2,\C) potential algebraic approach to the
construction of non-Hermitian Hamiltonians with real spectra. To this end we will show that
our potentials (\ref{eq:V^{(+)}_R})--(\ref{eq:V^{(-)}_I}) fit into such a scheme of complex
potentials. In the next section, we therefore make a few remarks about the
realization of the potential algebra sl(2,\C).\par
%
%==============================================================
%
\section{\boldmath sl(2,\mbox{\sixteenof C}) Potential Algebra}

\setcounter{equation}{0}

In Ref.~\cite{bagchi00c}, we made a detailed study of the sl(2,\C) algebra. Its underlying
commutation relations are 
\begin{equation}
  \left[J_0, J_{\pm}\right] = \pm J_{\pm}, \qquad \left[J_+, J_-\right] = - 2J_0.
\end{equation}
The generators $J_0$ and $J_{\pm}$ can be given a differential realization
\begin{equation}
  J_0 = - {\rm i} \frac{\partial}{\partial\phi}, \qquad J_{\pm} = e^{\pm{\rm i}\phi}
  \Biggl[\pm\frac{\partial}{\partial x} + \Biggl({\rm i} \frac{\partial}{\partial\phi}
  \mp \frac{1}{2}\Biggr) F(x) + G(x)\Biggr], 
\end{equation}
where the auxiliary variable $\phi \in [0, 2\pi)$ facilitates their closure and the two
functions $F(x)$, $G(x) \in \C$ are subjected to constraints of the form
\begin{equation}
  \frac{dF}{dx} = 1 - F^2, \qquad \frac{dG}{dx} = - FG, \qquad x \in \R. 
  \label{eq:constraints}
\end{equation}
Note that we have here $J_- \ne J_+^{\dagger}$, thereby inducing an sl(2,\C) algebra
rather than so(2,1), which is consistent with $J_- = J_+^{\dagger}$.\par
%
%-----------------------------------------------------------------------
%
In the case of either sl(2,\C) or so(2,1), the irreducible representations are furnished
by~\cite{englefield}
\begin{eqnarray}
  J_0 |km\rangle & = & m |km\rangle, \qquad m=k, k+1, k+2, \ldots, \nonumber \\
  J^2 |km\rangle & = & k(k-1) |km\rangle,  
\end{eqnarray}
which are essentially of the type D$^+_k$. The Casimir operator $J^2$ corresponds to
\begin{eqnarray}
  J^2 & = & J_0^2 \mp J_0 - J_{\pm} J_{\mp} \nonumber \\
  & = & \frac{\partial^2}{\partial x^2} - \left(\frac{\partial^2}{\partial \phi^2} +
  \frac{1}{4}\right) F' + 2 {\rm i} \frac{\partial}{\partial \phi} G' - G^2 - \frac{1}{4}.
\end{eqnarray}
\par
%
%-----------------------------------------------------------------------------------------------------------
%
Looking for representations that are
\begin{equation}
  |km\rangle = \Psi_{km}(x, \phi) = \psi_{km}(x) \frac{e^{{\rm i}
  m\phi}}{\sqrt{2\pi}}, 
\end{equation}
where $k > 0$ and $m = k+n$, $n=0$, 1, 2,~\ldots, we find $\psi_{km}(x)$ to satisfy the
Schr\"odinger equation
\begin{equation}
  - \psi''_{km} + V_m \psi_{km} = - \left(k - \case{1}{2}\right)^2 \psi_{km}.
  \label{eq:SE} 
\end{equation}
In (\ref{eq:SE}), the one-parameter family of potentials $V_m$ is given by
\begin{eqnarray}
  V_m  & = & - \left(m - \case{1}{2}\right) \left(m + \case{1}{2}\right) F' + 2m G' +
          G^2 \nonumber \\
  & = &  - \left(m - \case{1}{2}\right) \left(m + \case{1}{2}\right) \left(1 - F^2\right) -
          2m FG + G^2,   \label{eq:Vm}
\end{eqnarray}
where Eq.~(\ref{eq:constraints}) has been used. These potentials share the same real
energy eigenvalues 
\begin{equation}
  E^{(m)}_n = - \left(m - n - \case{1}{2}\right)^2,  \label{eq:Emn}  
\end{equation}
thus producing a potential algebra.\par
%
%-------------------------------------------------------------------------------------------------------------
%
Equations~(\ref{eq:constraints}) can be solved for the functions $F$ and $G$ to get a
quite complete realization of the sl(2,\C) algebra. The results obtained by us may be
summarized as follows:
\begin{equation}
\begin{array}{lll}
      {\rm I}: & F(x) = \tanh(x-c-{\rm i}\gamma), & G(x) = b \sech(x-c-{\rm i}\gamma),
            \\[0.2cm]
      {\rm II}: & F(x) = \coth(x-c-{\rm i}\gamma), & G(x) = b \cosech(x-c-{\rm i}\gamma),
            \\[0.2cm]
      {\rm III}: & F(x) = \pm 1, & G(x) = b e^{\mp x},  
\end{array}  \label{eq:FG}  
\end{equation}
where $b = b_R + {\rm i} b_I$, $b_R$, $b_I \in \R$, and $-\frac{\pi}{4} \le \gamma <
\frac{\pi}{4}$. These lead to potentials
\begin{eqnarray}
  {\rm I:\quad} V_m & = & \left[(b_R + {\rm i}b_I)^2 - m^2 + \case{1}{4}\right]
         \sech^2(x - c - {\rm i}\gamma) \nonumber \\ 
  && \mbox{} - 2m (b_R + {\rm i}b_I) \sech(x - c - {\rm i}\gamma)
         \tanh(x - c - {\rm i}\gamma), \\
  {\rm II:\quad} V_m & = & \left[(b_R + {\rm i}b_I)^2 + m^2 - \case{1}{4}\right]
         \cosech^2(x - c - {\rm i}\gamma) \nonumber \\
  && \mbox{} - 2m (b_R + {\rm i}b_I) \cosech(x - c - {\rm i}\gamma)
         \coth(x - c - {\rm i}\gamma), \\         
  {\rm III:\quad} V_m & = & (b_R + {\rm i}b_I)^2 e^{\mp 2x} \mp 2m (b_R + {\rm
         i}b_I) e^{\mp x}.  
\end{eqnarray}
\par
%
%-----------------------------------------------------------------------------------------------------------
%
In this way the group theoretical approach of the potential algebras can be extended to
non-Hermitian Hamiltonians (a subclass of which forms the PT-symmetric ones) by a
simple complexification of the real algebras considered for Hermitian Hamiltonians.\par
%
%=============================================================
%
\section{\boldmath sl(2,\mbox{\sixteenof C}) Potentials in SUSYQM}

\setcounter{equation}{0}

To realize sl(2,\C) potentials from the supersymmetry-inspired
Eqs.~(\ref{eq:V^{(+)}_R})--(\ref{eq:V^{(-)}_I}), we notice that $V_m$ can be considered
as a special case of the complex potential $V^{(+)} = W^2 - W' + E$ given
by~(\ref{eq:Vpm}), corresponding to the choice of the complex superpotential
\begin{equation}
  W = \left(m - \case{1}{2}\right) F - G,  \label{eq:complexW}
\end{equation}
and the real energy
\begin{equation}
E = E_R = - \left(m - \case{1}{2}\right)^2.  \label{eq:E_R}
\end{equation}
Inserting~(\ref{eq:complexW}) and~(\ref{eq:E_R}) into the definition of $V^{(+)}$, we
indeed get
\begin{eqnarray}
  V^{(+)} \equiv V_m & = & \left[\left(m - \case{1}{2}\right) F - G\right]^2 -
         \left[\left(m - \case{1}{2}\right) \left(1 - F^2\right) + FG\right] - \left(m -
         \case{1}{2}\right)^2 \nonumber \\
  & = & - \left(m - \case{1}{2}\right) \left(m + \case{1}{2}\right) \left(1 - F^2\right) -
         2mFG + G^2,
\end{eqnarray}
which coincides with the sl(2,\C) form~(\ref{eq:Vm}).\par
%
%-------------------------------------------------------------------------------------------------------------
%
The potential $V^{(-)} = W^2 + W' + E$ of the superpartner is now
\begin{eqnarray}
  V^{(-)} & = & \left[\left(m - \case{1}{2}\right) F - G\right]^2 +
         \left[\left(m - \case{1}{2}\right) \left(1 - F^2\right) + FG\right] - \left(m -
         \case{1}{2}\right)^2 \nonumber \\
  & = & - \left(m - \case{3}{2}\right) \left(m - \case{1}{2}\right) \left(1 - F^2\right) -
         2 (m-1) FG + G^2 \nonumber \\
  & = & V_{m-1}.  \label{eq:Vm-1}
\end{eqnarray}
From~(\ref{eq:Emn}), it is obvious that we are here in the case where $H_-$ has one
level less than $H_+$ and Eq.~(\ref{eq:gs+}) applies.\par
%
%------------------------------------------------------------------------------------------------------------
%
In Table~1 we have displayed the various forms of the complex superpotential $W$ for
different solutions of $F$ and $G$ summarized in~(\ref{eq:FG}).\par
%
%----------------------------------------------------------------------------------------------------------
%
It is interesting to discuss the results corresponding to the choice $\gamma = 0$. While
for $b_I = 0$, the superpotential along with the partner potentials reduce to their real
forms which is only expected, the possibility $b_R = 0$ is worth taking a look. For case~I,
$W$ simply becomes
\begin{equation}
  W = \left(m - \case{1}{2}\right) \tanh (x-c) - {\rm i} b_I \sech (x-c),
\end{equation}
leading to
\begin{equation}
  V^{(+)} \equiv V_m = \left(- b_I^2 - m^2 + \case{1}{4}\right) \sech^2 (x-c) - 2 {\rm i}
  m b_I \sech (x-c) \tanh (x-c).
\end{equation}
Its superpartner can be read off readily from~(\ref{eq:Vm-1}) and is
\begin{equation}
  V^{(-)} \equiv V_{m-1} = \left[- b_I^2 - (m-1)^2 + \case{1}{4}\right] \sech^2 (x-c) - 2
  {\rm i} (m-1) b_I \sech (x-c) \tanh (x-c).
\end{equation}
Note that both $V^{(+)}$ and $V^{(-)}$ turn out to be PT symmetric. For completeness
we give the solutions for $f$, $g$, $V^{(\pm)}_R$, and $V^{(\pm)}_I$. These are
\begin{eqnarray}
  f & = & \left(m - \case{1}{2}\right) \tanh (x-c),  \label{eq:spe-f} \\
  g & = & - b_I \sech (x-c), \\
  V^{(+)}_R & = & \left(- b_I^2 - m^2 + \case{1}{4}\right) \sech^2 (x-c), 
          \label{eq:spe-V^{(+)}_R} \\
  V^{(+)}_I & = & - 2 m b_I \sech (x-c) \tanh (x-c),  \label{eq:spe-V^{(+)}_I} \\
  V^{(-)}_R & = & \left[- b_I^2 - (m-1)^2 + \case{1}{4}\right] \sech^2 (x-c), 
                    \label{eq:spe-V^{(-)}_R} \\
  V^{(-)}_I & = &  - 2 (m-1) b_I \sech (x-c) \tanh (x-c)  \label{eq:spe-V^{(-)}_I}.    
\end{eqnarray}
\par
%
%--------------------------------------------------------------------------------------------------------------
%
A particular case of the above scheme corresponding to $m=1$ was derived by Bagchi
and Roychoudhury~\cite{bagchi00a}, who showed that the PT-symmetric combination
of~(\ref{eq:spe-V^{(+)}_R}) and~(\ref{eq:spe-V^{(+)}_I}) has energy levels that are
negative semi-definite and, except for the zero-energy state, coincide with those
of the $\sech^2$ potential resulting from~(\ref{eq:spe-V^{(-)}_R}).\par
%
%---------------------------------------------------------------------------------------------------------
%
Similarly we can deal with cases~II and~III, both of which are PT non-symmetric, for the
choice of parameters $\gamma = 0$ and $b_R = 0$. While case~II gives the $\coth$,
$\cosech$ version of (\ref{eq:spe-f})--(\ref{eq:spe-V^{(-)}_I}), case~III leads to the
complexified Morse potential. The results for $W$, $f$, and $g$ are
\par
\noindent
Case~II for $\gamma = 0$, $b_R = 0$:
\begin{eqnarray}
  W & = & \left(m - \case{1}{2}\right) \coth (x-c) - {\rm i} b_I \cosech (x-c),
             \nonumber \\
  f & = & \left(m - \case{1}{2}\right) \coth (x-c), \nonumber  \\
  g  & = & - b_I \cosech (x-c), 
\end{eqnarray}
Case~III for $\gamma = 0$, $b_R = 0$:
\begin{eqnarray}
  W & = & \pm \left(m - \case{1}{2}\right) - {\rm i} b_I e^{\mp x}, \nonumber \\
  f & = & \pm \left(m - \case{1}{2}\right), \nonumber \\
  g & = & - b_I e^{\mp x}. 
\end{eqnarray}
\par
%
%============================================================
%
\section{\boldmath A PT-Symmetric Potential in Terms of Weierstrass $\wp$ Function}

\setcounter{equation}{0}

In this section we make a specific attempt to search for a PT-symmetric potential
described by Weierstrass $\wp$ function. In Ref.~\cite{andrianov}, Andrianov {\em et al.}
took $V^{(+)}_R = 0$, $V^{(+)}_I = 0$ to analyze complex transparent potentials
belonging to the set (\ref{eq:V^{(+)}_R})--(\ref{eq:V^{(-)}_I}). Here we consider an
equally viable possibility by setting $V^{(-)}_R = 0$, $V^{(+)}_I = 0$. This case is
indeed nontrivial since other possibilities are either related to Andrianov {\em et al.}
conjecture or the present one under $f \to -f$ and $g \to -g$.\par 
% 
%----------------------------------------------------------------------------------------------------------
% 
While $V^{(+)}_I = 0$ results in
\begin{equation}
  f = \frac{g'}{2g},  \label{eq:f}
\end{equation}
leading to
\begin{eqnarray}
  V^{(+)}_R & = & \frac{3g^{\prime 2}}{4g^2} - \frac{g''}{2g} - g^2 + E_R, \\
  V^{(-)}_R & = & - \frac{g^{\prime 2}}{4g^2} + \frac{g''}{2g} - g^2 + E_R, \\
  V^{(-)}_I & = & 2g',
\end{eqnarray}
$V^{(-)}_R = 0$ gives us the solution for $g$ in terms of the differential equation
\begin{equation}
  \frac{dg}{\sqrt{g \left(\frac{4}{3} g^3 - 4E_R g + a\right)}} = \pm dx,  \label{eq:diff}
\end{equation}
where $a$ represents a constant of integration, which we take to be non-zero.\par
%
%----------------------------------------------------------------------------------------------------------
%
Writing $y(g) = g \left(\frac{4}{3} g^3 - 4E_R g + a\right)$ in the form $y(g) = a_0
g^4 + 4a_1 g^3 + 6a_2 g^2 + 4a_3 g + a_4$, we have $a_0 = \frac{4}{3}$, $a_1 =
0$, $a_2 = - \frac{2}{3} E_R$, $a_3 = \frac{1}{4} a$, $a_4 = 0$. We next define
quantities $g_2$ and $g_3$ as
\begin{eqnarray}
  g_2 & = & a_0 a_4 - 4a_1 a_3 + 3a_2^2 = \case{4}{3} E_R^2, \\
  g_3 & = & a_0 a_2 a_4 + 2a_1 a_2 a_3 - a_2^3 - a_0 a_3^2 - a_1^2 a_4 =
\case{8}{27} E_R^3 - \case{1}{12} a^2.
\end{eqnarray}   
\par
%
%----------------------------------------------------------------------------------------------------------
% 
Let
\begin{equation}
  z = \int_{g_0}^g \, [y(t)]^{-1/2} dt,  \label{eq:z}
\end{equation}
where $g_0$ is any root of the equation $y(g) = 0$. We identify $g_0$ as $g_0 =
0$.\par
%
%---------------------------------------------------------------------------------------------------------
%
Let us substitute $t = \frac{1}{\tau}$ and $g = \frac{1}{\xi}$ to rewrite~(\ref{eq:z}) as 
\begin{equation}
  z = \int_{\xi}^{\infty} \, \left(4a_3 \tau^3 + 6a_2 \tau^2 + a_0\right)^{-1/2} d\tau.
\end{equation}
The second term in the integrand can be removed by effecting the transformations
\begin{eqnarray}
  \tau & = & \frac{\sigma - \frac{1}{2} a_2}{a_3} = \frac{4}{a} \left(\sigma +
         \frac{1}{3} E_R\right), \\
  \xi & = & \frac{s - \frac{1}{2} a_2}{a_3} = \frac{4}{a} \left(s + \frac{1}{3} E_R\right).
         \label{eq:xi}
\end{eqnarray}
As a result, $z$ turns out to be given by
\begin{equation}
  z = \int_s^{\infty} \, \left(4 \sigma^3 - g_2 \sigma - g_3\right)^{-1/2} d\sigma.
\end{equation}
\par
%
%------------------------------------------------------------------------------------------------------------
%
Now from the theory of elliptic functions~\cite{whittaker}, we can read off
\begin{equation}
  s = \wp(z; g_2, g_3),
\end{equation}
where $\wp(z; g_2, g_3)$ is Weierstrass elliptic function with $g_2$ and $g_3$ as
invariants. Equation~(\ref{eq:xi}) yields
\begin{equation}
  g = \frac{a}{4} \left[\wp(z; g_2, g_3) + \frac{1}{3} E_R\right]^{-1}.  \label{eq:g}
\end{equation}
\par
%
%---------------------------------------------------------------------------------------------------------
%
Hence we deduce
\begin{eqnarray}
  V^{(-)}_I & = & - \frac{a}{2} \frac{\wp'(z)}{\left[\wp(z) + \frac{1}{3} E_R\right]^2},
        \label{eq:ellipV-}\\
  V^{(+)}_R & = & 2 \left[\wp(z) + \frac{1}{3} E_R\right] -  \frac{a^2}{12 \left[\wp(z) +
        \frac{1}{3} E_R\right]^2}.  \label{eq:ellipV+}
\end{eqnarray}
In deriving~(\ref{eq:ellipV+}), use has been made of the differential equation satisfied by
$\wp(z)$,
\begin{equation}
  \wp^{\prime2}(z) = 4 \wp^3(z) - g_2 \wp(z) - g_3,
\end{equation}
and of its consequence
\begin{equation}
  \wp''(z) = 6 \wp^2(z) - \case{1}{2} g_2.
\end{equation}
We notice that while $V^{(+)} = V^{(+)}_R$ is pure real, $V^{(-)} = {\rm i} V^{(-)}_I$ is
a pure imaginary potential.\par
%
%--------------------------------------------------------------------------------------------------------------
%
Comparing (\ref{eq:diff}) with~(\ref{eq:z}), we see that $z = \pm x + c$, where $c$ is
an integration constant. Since $\wp(z)$ and $\wp'(z)$ are respectively even and odd
functions of~$z$, the two solutions obtained by taking either sign correspond to each
other by a mere change of signs of the integration constants $a$ and $c$. Thus without
any loss of generality we can take
$z = x + c$ only.\par
%
%---------------------------------------------------------------------------------------------------------
%
We may distinguish between the non-degenerate and degenerate cases of Weierstrass
$\wp$ function~\cite{whittaker}.\par
%
%------------------------------------------------------------------------------------------------------------
%
In the non-degenerate case, the roots of the cubic equation
\begin{equation}
  4 \sigma^3 - g_2 \sigma - g_3 = 0  \label{eq:cubic}
\end{equation}
are all distinct and the corresponding discriminant 
\begin{equation}
  D = g_2^3 - 27 g_3^2 = \frac{a^2}{48} \left(64 E_R^3 - 9a^2\right)
\end{equation}
is non-vanishing. It is either positive or negative according to whether $|a| < \frac{8}{3}
E_R^{3/2}$ or $|a| > \frac{8}{3} E_R^{3/2}$, the former case occurring only for $E_R >
0$.\par
%
%--------------------------------------------------------------------------------------------------------
%
By using numerical studies, we showed that in the $D < 0$ case, wherein the Weierstrass
function has a pair of complex conjugate primitive periods $2 \omega$, $2\omega' = 2
\omega^*$, $V^{(+)}_R$ and $V^{(-)}_I$ go to $- \infty$ for some $z$ values because
the denominators in~(\ref{eq:ellipV-}) and~(\ref{eq:ellipV+}) vanish for such values. On
the contrary, in the $D > 0$ case, wherein the Weierstrass function has a pair of primitive
periods $2 \omega$, $2\omega'$ with $\omega$ real and $\omega'$ imaginary, we
obtain well-behaved potentials defined on the interval $0 < z < 2\omega$ or $- c < x
< 2\omega -c$ . The potential $V^{(+)}_R$ is a single-well potential, singular at $z \to 0$
and $z \to 2\omega$ (where it behaves as $1/z^2$ and $1/(z - 2\omega)^2$,
respectively), and symmetric around its minimum at $z = \omega$, whereas $V^{(-)}_I$
vanishes at $z \to 0$ and $z \to 2\omega$, and is antisymmetric around $z = \omega$.
Hence, the potential $V^{(-)} = {\rm i} V^{(-)}_I$ is PT symmetric provided the parity
operation is defined with respect to a mirror placed at $z = \omega$ or $x = \omega -
c$.\par
%
%---------------------------------------------------------------------------------------------------------
%
In Fig.~1, the functions $V^{(+)}_R$ and $V^{(-)}_I$ are displayed in terms of $z$ for
$E_R = \sqrt{3}$ and $a = 4 \sqrt{2/\sqrt{3}}$, corresponding to the invariants $g_2 =
4$, $g_3 = 0$. For such values, the cubic equation~(\ref{eq:cubic}) has the three real
roots $e_1 = 1$, $e_2 = 0$, $e_3 = -1$, and the real primitive period is given by
$2\omega = 2 x_*$ where $x_* = \sqrt{\pi}\, \Gamma(5/4)/\Gamma(3/4) \simeq
1.311$~\cite{shifman}. The minimum of $V^{(+)}_R$ is equal to $6 \left(1 -
\frac{1}{\sqrt{3}}\right)$.\par
%
%----------------------------------------------------------------------------------------------------------
%  
In the degenerate case, at least two of the roots of~(\ref{eq:cubic}) are equal, meaning
that $D=0$. This condition imposes that  $E_R > 0$, $a = \pm \frac{8}{3} E_R^{3/2}$,
$g_2 = \frac{4}{3} E_R^2$, and $g_3 = - \frac{8}{27} E_R^3$. We are then in a case
where the real period becomes infinite and $\wp(z)$ reduces to~\cite{whittaker}
\begin{equation}
  \wp(z) = E_R \left[\case{1}{3} + \cosech^2\left(\sqrt{E_R}\, z\right)\right].
  \label{eq:P-deg}
\end{equation}
\par
%
%----------------------------------------------------------------------------------------------------------
%
In consequence we have
\begin{equation}
  g = \pm \frac{\sqrt{E_R}}{1 + \frac{3}{2} \cosech^2\left(\sqrt{E_R}\, z\right)},
\end{equation}
from which we obtain
\begin{eqnarray}
  V^{(+)}_R & = & \frac{4}{3} E_R \left\{1 + \frac{3}{2} \cosech^2\left(\sqrt{E_R}\,
        z\right) - \frac{1}{\left[1 + \frac{3}{2} \cosech^2\left(\sqrt{E_R}\,
        z\right)\right]^2}\right\},  \\
  V^{(-)}_I & = & \pm 6E_R \frac{\cosech^2\left(\sqrt{E_R}\, z\right)
        \coth\left(\sqrt{E_R}\, z\right)} {\left[1 + \frac{3}{2} \cosech^2\left(\sqrt{E_R}\,
        z\right)\right]^2},
\end{eqnarray}
defined on the interval $0 < z < \infty$ or $- c < x < \infty$. As such $V^{(-)}$, given by
$V^{(-)} = {\rm i} V^{(-)}_I$, is a non-PT-symmetric potential. We also
note that  as $z \to 0$, $V^{(+)}_R  \sim 2/z^2$ while for $z \to \infty$, $V^{(+)}_R
\sim 24 E_R \exp(-2 \sqrt{E_R}\, z)$. These are reasonable boundary conditions, the
behaviour of $V^{(+)}_R$ proving to be repulsive.\par
%
%---------------------------------------------------------------------------------------------------------
%
An example is displayed in Fig.~2 for $E_R = \sqrt{3}$ and $a = 8/3^{1/4}$,
corresponding to the invariants $g_2 = 4$ and $g_3 = - 8/3^{3/2}$. For such values, the
cubic equation~(\ref{eq:cubic}) has the three real roots $e_1 = e_2 = 1/\sqrt{3}$ and
$e_3 = - 2/\sqrt{3}$.\par
%
%--------------------------------------------------------------------------------------------------------
%
As discussed in Sec.~2, the spectra of the supersymmetric partners $H_+$ and $H_-$ are
alike except for the ground state. In the non-degenerate case, whenever $E_R > 0$ and
$|a| < \frac{8}{3} E_R^{3/2}$, the Hermitian Hamiltonian $H_+$ has an infinite number of
(unknown) discrete positive-energy levels. The same is true for the PT-symmetric
Hamiltonian $H_-$, but in addition the latter has a normalizable eigenfunction
$\psi^{(-)}_0$ corresponding to $E=0$. From (\ref{eq:gs-}), (\ref{eq:W}),
and~(\ref{eq:f}), it is given by
\begin{equation}
  \psi^{(-)}_0(x) = K \sqrt{g} \exp\left({\rm i} \int^x g(t) dt\right).  \label{eq:psi-add}    
\end{equation}
By using~(\ref{eq:g}), its modulus can be written as
\begin{equation}
  |\psi^{(-)}_0(x)| = \frac{|K| \sqrt{|a|}}{2} \left[\wp(z;g_2,g_3) +
  \frac{1}{3} E_R\right]^{-1/2}.  
\end{equation}
Hence it vanishes at $z = 0$ and $z = 2\omega$, and is regular in between, showing that
$\psi^{(-)}_0(x)$ is indeed normalizable on $(-c, 2\omega -c)$.\par
%
%------------------------------------------------------------------------------------------------------------
%
Similarly, in the degenerate case, i.e., for $E_R > 0$ and $|a| = \frac{8}{3} E_R^{3/2}$,
the Hamiltonians $H_+$ and $H_-$ have both a continuous spectrum of unbounded
positive-energy states. In addition, $H_-$ has an unbound zero-energy state, whose
wave function is still given by~(\ref{eq:psi-add}). From~(\ref{eq:g})
and~(\ref{eq:P-deg}), we indeed obtain
\begin{equation}
  |\psi^{(-)}_0(x)| = \frac{|K|}{2} \sqrt{\frac{|a|}{E_R}} \left[\cosech^2\left(\sqrt{E_R}\,
z\right) + \frac{2}{3}\right]^{-1/2},
\end{equation}
which vanishes for $z=0$ and goes to $\frac{|K|}{2} \sqrt{\frac{3|a|}{2E_R}}$ for $z
\to \infty$.\par
%
%------------------------------------------------------------------------------------------------------------
%
As a final point, it is worth noting that the known zero-energy eigenfunction of $H_-$ is
an eigenfunction of the Hamiltonian $- \frac{d^2}{dx^2} + V^{(-)}(x)$ with energy
$E_R$.\par
%
%============================================================
%  
\section{Conclusion}

In the present paper, we both constructed some new PT-preserving or non-PT-preserving
complex potentials and analyzed some known ones from a SUSYQM viewpoint extended to
deal with non-Hermitian Hamiltonians.\par
%
%----------------------------------------------------------------------------------------------------------
%
To start with, we considered three sets of complex potentials, recently derived by a
potential algebraic approach based on the complex Lie algebra sl(2,\C)~\cite{bagchi00c},
and proved that they can be generated as well from a complex superpotential  and a pair
of supercharges that are not related by Hermitian conjugation. This extends to the
complex domain the well-known relationship between SUSYQM and potential algebras for
Hermitian Hamiltonians, resulting from their common link~\cite{barut} with the
factorization method~\cite{infeld} and Darboux transformations~\cite{darboux}.\par
%
%------------------------------------------------------------------------------------------------------------
%
In a second step, we analyzed the special case of the extended SUSYQM
theory~\cite{andrianov} wherein the starting potential is real and its partner imaginary. 
This allowed us to build for the first time a pair of complex partner potentials defined in
terms of Weierstrass elliptic function. The PT-symmetric imaginary partner
potential has one known eigenvalue equal to the factorization energy $E_R$ and
corresponding to a bound state. When the Weierstrass function degenerates to a
hyperbolic one, the imaginary partner potential becomes PT non-symmetric and its
eigenvalue $E_R$ corresponds to an unbound state. We have therefore constructed two
new quasiexactly solvable complex potentials~\cite{bagchi00b, khare}.\par
%
%=============================================================
%
\section*{Acknowledgments}

One of us (S.\ M.) thanks the Council of Scientific and Industrial Research, New Delhi for
financial support.\par
%
%==============================================================
% 

%
%=============================================================
%
\newpage

\begin{table}[h]

\caption{The functions $F$, $G$, and the superpotential $W$ corresponding to the three
cases considered in the text.}

\vspace{1cm}
\begin{center}
\begin{tabular}{cccc}
  \hline\\[-0.2cm]
  Cases & $F$ & $G$ & $W$\\[0.2cm]
  \hline\\[-0.2cm]
  I & $\tanh(x - c - {\rm i} \gamma)$ & $b \sech(x - c - {\rm i} \gamma)$ & $\left(m -
        \case{1}{2}\right) \tanh(x - c - {\rm i} \gamma)$ \\[0.2cm]
  & & & $- b \sech(x - c - {\rm i} \gamma)$\\[0.2cm]
  II& $\coth(x - c - {\rm i} \gamma)$ & $b \cosech(x - c - {\rm i} \gamma)$ & $\left(m -
        \case{1}{2}\right) \coth(x - c - {\rm i} \gamma)$ \\[0.2cm]
  & & & $- b \cosech(x - c - {\rm i} \gamma)$\\[0.2cm]
  III & $\pm 1$ & $b e^{\mp x}$ & $\pm \left(m - \case{1}{2}\right) - b e^{\mp x}$
        \\[0.2cm]
  \hline
\end{tabular}
\end{center}

\end{table} 
%
%==============================================================
%
\newpage
\section*{Figure captions}

{}Fig.\ 1. (a) $V^{(+)}_R$ and (b) $V^{(-)}_I$ in terms of $z = x + c$ for the
non-degenerate case of Weierstrass $\wp$ function, $E_R = \sqrt{3}$, and $a = 4
\sqrt{2/\sqrt{3}}$.

\medskip\noindent
{}Fig.\ 2. (a) $V^{(+)}_R$ and (b) $V^{(-)}_I$ in terms of $z = x + c$ for the
degenerate case of Weierstrass $\wp$ function and $E_R = \sqrt{3}$.
\par
%
%======================================================================
% 
\newpage
\begin{picture}(160,100)
\put(35,0){\mbox{\scalebox{1.0}{\includegraphics{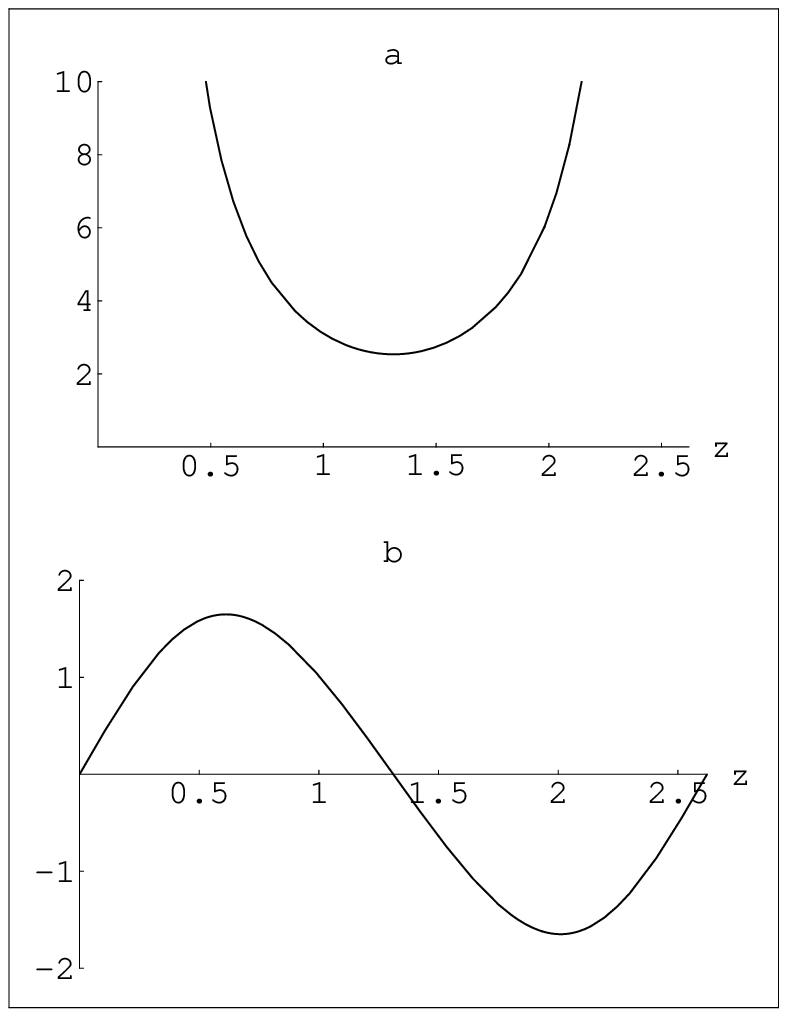}}}}
\end{picture}
\vspace{5cm}
\centerline{Figure 1}
%
%----------------------------------------------------------------------
%
\newpage
\begin{picture}(160,100)
\put(35,0){\mbox{\scalebox{1.0}{\includegraphics{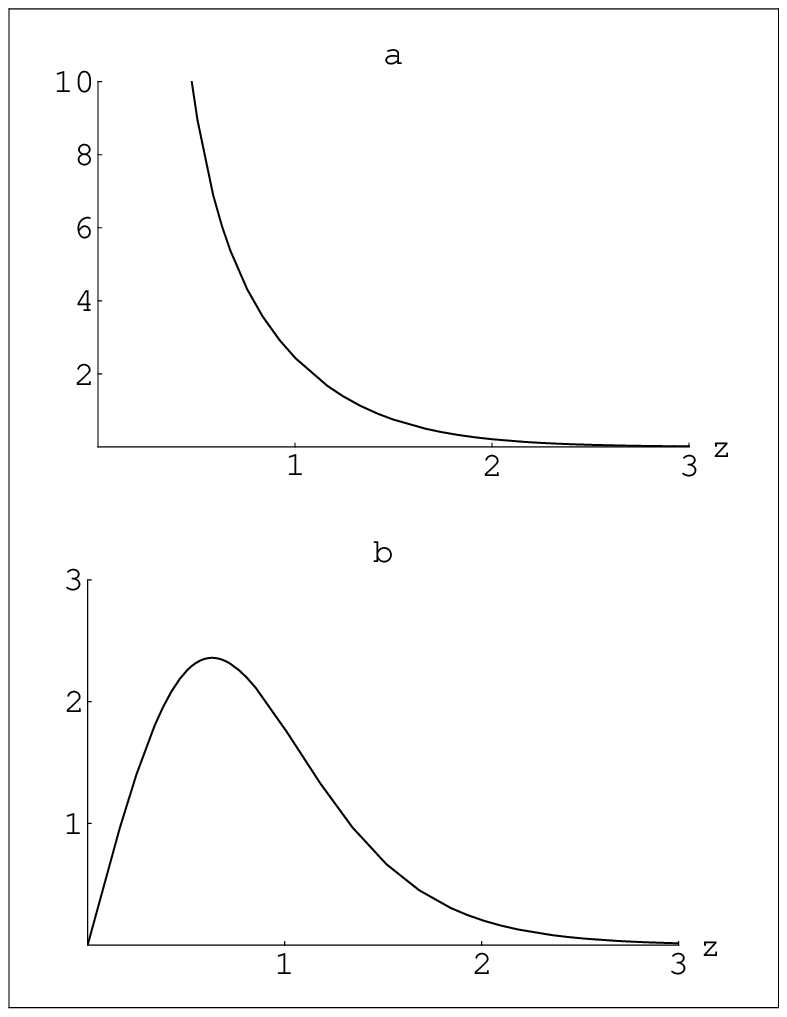}}}}
\end{picture}
\vspace{5cm}
\centerline{Figure 2}

\end{document}